\documentclass[%
reprint,
superscriptaddress,
amsmath,
amssymb,
aps,
prl,
longbibliography
]{revtex4-2}

\usepackage[T1]{fontenc}
\usepackage[utf8]{inputenc}

\usepackage{times} 
\usepackage[unicode=true,breaklinks=true,colorlinks=true]{hyperref}
\hypersetup{citecolor=blue,urlcolor=blue}
\usepackage{graphicx}
\usepackage{bm}
\usepackage{bbold}
\usepackage{braket}
\usepackage[dvipsnames]{xcolor}
\usepackage{tikz}
\usepackage{pdfpages} 

\usepackage{mathtools}
\usepackage{tensor}

\newcommand\norm[1]{\left\lVert#1\right\rVert}

\makeatletter 
\AtBeginDocument{\let\LS@rot\@undefined} 
\makeatother

\begin{document}
\title{Thermodynamic principle for quantum metrology}
\author{Yaoming Chu}
\affiliation{School of Physics, Institute for Quantum Science and Engineering, International Joint Laboratory on Quantum Sensing and Quantum Metrology,  Huazhong University of Science and Technology, Wuhan 430074, China}
\affiliation{Wuhan National High Magnetic Field Center, Wuhan National Laboratory for Optoelectronics, Huazhong University of Science and Technology, Wuhan 430074, China}
\author{Jianming Cai}
\email{jianmingcai@hust.edu.cn}
\affiliation{School of Physics, Institute for Quantum Science and Engineering, International Joint Laboratory on Quantum Sensing and Quantum Metrology,  Huazhong University of Science and Technology, Wuhan 430074, China}
\affiliation{Wuhan National High Magnetic Field Center, Wuhan National Laboratory for Optoelectronics, Huazhong University of Science and Technology, Wuhan 430074, China}

\begin{abstract}
The heat dissipation in quantum metrology represents not only an unavoidable problem towards practical applications of quantum sensing devices but also a fundamental relationship between thermodynamics and quantum metrology. However, a general thermodynamic principle which governs the rule of energy consumption in quantum metrology, similar to Landauer's principle for heat dissipation in computations, has remained elusive. Here, we establish such a physical principle for energy consumption in order to achieve a certain level of measurement precision in quantum metrology, and show that it is intrinsically determined by the erasure of quantum Fisher information. The principle provides a powerful tool to investigate the advantage of quantum resources, not only in measurement precision but also in energy efficiency. It also serves as a bridge between thermodynamics and various fundamental physical concepts related in quantum physics and quantum information theory.
\end{abstract}

\maketitle

{\it Introduction.---}Deeper understanding of thermodynamic energy cost for information processing, and further reducing the heat generated by computations have always been a fundamental goal in information technology \cite{Bennett1982}. The ultimate physical limit on energy consumption of computation is set by Landauer's principle \cite{Landauer1961,Shizume1995,Piechocinska2000,Rio2011,Goold2015,Lorenzo2015,Proesmans2020,Miller2020,Toyabe2010,Berut2012,Peterson2016,Yan2018}. It states that the irreversible erasure of information would inevitably dissipate a certain amount of heat into environment, while reversible operations can in principle be implemented at no energy cost. Landauer's principle establishes a fundamental relationship between information theory and thermodynamics \cite{Plenio2001,Strasberg2017}. It represents the philosophy of ``{\it information is physical} '' \cite{Landauer1991}, and is the key to the exorcism of Maxwell's Demon \cite{Bennett2003,Sagawa2009,Maruyama2009}.

Quantum metrology, as a fast-developing field of quantum technology, achieves highly sensitive measurements of physical parameters over classical techniques \cite{Giovannetti2011,Degen2017,Pezze2018,Braun2018}. Understanding the quantum limits of quantum metrology, e.g., in terms of measurement precision and channel capacity, has yielded fundamental insights into its connections to quantum geometry \cite{Braunstein1994}, many-body entanglement \cite{Sorensen2001}, and Shannon-Hartley theorem \cite{Sidles2009}. Similar to computation, the energy consumption in quantum metrology would become an important issue towards potentially massive applications of 
quantum sensing devices, and represents a fundamental link between thermodynamics and quantum metrology. So, is there a general principle that determines the limit of energy cost in quantum metrology? The analysis of certain examples hints that work cost might be relevant in quantum metrology \cite{Erker2017,Lipka2018,Liuzzo2018}. But these attempts are confined to specific models, and do not provide a complete answer. Here we take a very different approach and establish a fundamental principle for heat dissipation in quantum metrology.

As our main result, we find the physical limit on energy consumption of quantum metrology, and demonstrate that it essentially arises from the erasure of quantum Fisher information (QFI) which determines the best achievable measurement precision \cite{Braunstein1994}. The result establishes a basic thermodynamic principle for quantum metrology, and clearly states the thermodynamic cost in order to achieve a certain level of measurement precision. We provide an efficient way to investigate energy efficiency of multi-qubit states for quantum metrology, and point out that it is possible to achieve Heisenberg limit of measurement precision with energy consumption that does not increase with the number of probes. Furthermore, the QFI plays important roles in various fundamental quantum phenomena, such as multipartite entanglement \cite{Pezze2009,Hyllus2012,Strobel2014,Hauke2016}, quantum criticality \cite{Zanardi2007} and quantum geometry \cite{Braunstein1994}. Therefore, the present thermodynamics of quantum metrology would inspire us to explore the thermodynamic meaning of the related physical concepts involved in these phenomena.

\begin{figure*}
\centering
\includegraphics[width=12cm]{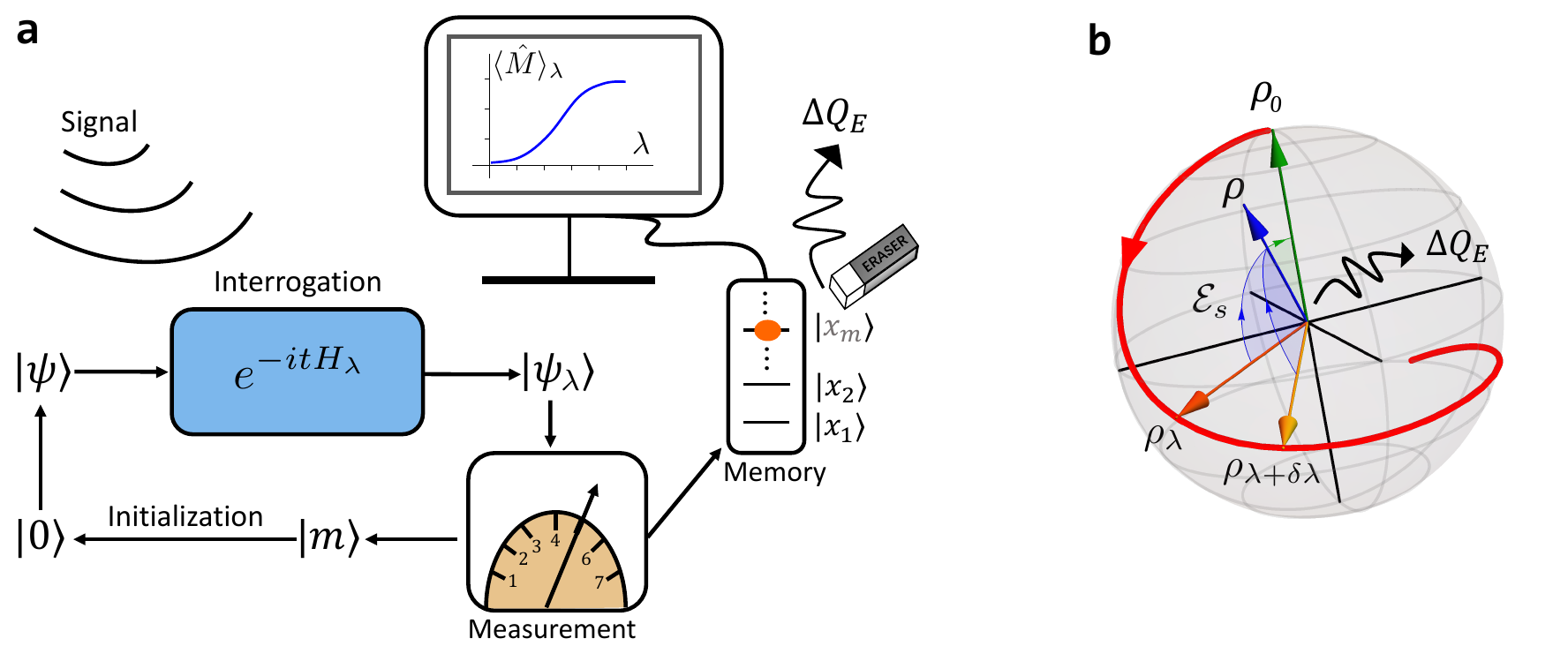}
\caption{\textbf{Heat dissipation in quantum metrology.} \textbf{a,} In standard quantum metrology, after the interrogation for time $t$, the parametrized final state $\vert \psi_\lambda\rangle$ is measured, which provides information on the unknown parameter $\lambda$. To realize a closed metrological cycle, both the quantum probe system and the memory need to be recovered to their initial states. A unitary operation can evolve the system into its initial state $|0\rangle$ without heat dissipation; while the erasure of information stored in the memory about the measurement outcomes inevitably results in heat dissipation. \textbf{b,} The QFI characterizes the distinguishability between neighboring parametrized quantum states $\rho_\lambda$ and $\rho_{\lambda+\delta\lambda}$, which is erased as they are transformed into a parameter-independent state $\rho$ via the map $\mathcal{E}_s$ (blue line). The erasure of QFI during this process will cause heat dissipation or entropy increase, see Eq.\,\eqref{eq:Q-S-F}. A possible subsequent operation (green line) that transforms $\rho$ back to the initial pure state may dissipate additional heat into environment according to Landauer's principle. }\label{fig:model}
\end{figure*}

{\it Background on quantum metrology.---}In quantum metrology, one generally prepares a quantum probe system in an initial state $\rho_0=\vert{\psi}\rangle \langle{\psi}\vert$, which subsequently undergoes a parameter-dependent evolution $U_{\lambda}$ to the final state $\rho_{\lambda}\equiv|\psi_\lambda\rangle\langle\psi_\lambda|=U_{\lambda}\rho_0 U_{\lambda}^{\dagger}$ \cite{Boixo2007,Toth2014,Liu2015,Pang2017}. In order to access information about the parameter, quantum metrology requires to perform measurement on the parametrized state $\rho_{\lambda}$. Such a measurement in the basis of $\{|m\rangle\langle m|\}_{m=1}^d$ ($d$ is the dimension of the system) can be described as \cite{Sagawa2009,Zurek1981}
\begin{equation}
  \label{eq:M}
  \mathcal{M}:\rho_\lambda \otimes |x\rangle\langle x| \to \sum_m \alpha_m^\lambda |m\rangle\langle m|\otimes |x_m\rangle\langle x_m|,
\end{equation}
where $\alpha_m^\lambda=\langle m|\rho_\lambda|m\rangle$, and $\{|x_m\rangle\}$ represents the internal structure of a memory, interacting with the system and storing the measurement outcomes, see Fig.\,\ref{fig:model}a. The ultimate precision of parameter estimation, achieved by performing optimal measurements on $\rho_{\lambda}$, is determined by the QFI,
\begin{equation}
F_Q[\rho_\lambda]=4\left[\langle \partial_\lambda\psi_\lambda|\partial_\lambda\psi_\lambda\rangle - \left|\langle \psi|\partial_\lambda\psi_\lambda\rangle\right|^2\right],
\end{equation}
 that represents a crucial measure of how much information a parametrized quantum state contains about the unknown parameter \cite{Braunstein1994}. 
 

%

{\it Main result.---}In the standard framework of quantum metrology, see Fig.\ref{fig:model}a, the parameter information is transferred to the memory after the measurement. A unitary operation brings the system back to the initial state without heat dissipation into environment because of its reversibility \cite{Peres1985}. In order to realize a closed metrological cycle, the memory must be recovered to its original standard state $|x\rangle$, leading to a heat dissipation $\Delta Q_E^\lambda$ into environment according to Landauer's principle. Given an optimal measurement basis to achieve the best precision of parameter estimation, the Shannon entropy,  $\mathcal{S}$, which quantifies the amount of information associated with the measurement outcomes, is found to be determined by the QFI  \cite{supplement}, namely
\begin{equation}
  \label{eq:S-Fq}
  \mathcal{S}\geq \log (2) \norm{h_\lambda}^{-2} F_Q[\rho_\lambda],
\end{equation}
where $h_\lambda=i U_\lambda^\dagger \partial_\lambda U_\lambda$ represents the local generator of parametric translation, and $\norm{h_\lambda}$ denotes the operator seminorm \cite{Boixo2007}. The result in Eq.\,\eqref{eq:S-Fq} relates two information-theoretic quantities of particular interest in different fields, i.e., the QFI in quantum metrology and the Shannon entropy in information theory. It opens the possibility to establishes a relationship connecting quantum metrology and thermodynamics, as illustrated by the following theorem \cite{supplement}.


\noindent {\it\bf Theorem 1}
Given an environment of temperature $T$, the heat dissipation to extract the parameter information via an optimal measurement protocol on the parametrized state $\rho_\lambda = U_\lambda \rho_0 U_\lambda^\dagger$ (with $\rho_0=\vert\psi\rangle \langle \psi\vert$) is lower bounded by
\begin{equation}
    \label{eq:Q-Fq-lambda}
    \Delta Q_E^\lambda \geq \log (2) k_B T \norm{h_\lambda}^{-2} F_Q[\rho_\lambda],
\end{equation}
where $k_B$ is the Boltzmann constant, and $h_\lambda$ is the local generator of parametric translation.

Theorem 1 straightforwardly leads to some important implications in quantum metrology. In the traditional scenario of quantum metrology, one typically considers interrogation Hamiltonians of the form \cite{Boixo2007}, which have been implemented in a number of systems \cite{Giovannetti2011, Degen2017,Pezze2018,Braun2018},
\begin{equation}
H_{\lambda}=\lambda h.
\label{eq:trad-H}
\end{equation}
The local generator of the above Hamiltonian is given by $h_\lambda=ht$ with $t$ the interrogation time. Without loss of generality, we assume a normalized energy scale by requiring the operator seminorm $\norm{h}=1$ \cite{Boixo2007}. Stated by the so-called quantum Cram\'er-Rao bound (CRB) \cite{Braunstein1994}, the measurement precision as quantified by the variance of unbiased estimators \cite{Kay1993,Demkowicz2015} for the parametrized state $\rho_\lambda$ is lower bounded by $(\delta\lambda)^{2}\geq F_Q^{-1}[\rho_\lambda]$. According to Theorem 1, we find that the ultimate measurement precision is limited by the interrogation time $t$ and heat dissipation $ \Delta Q_E^\lambda$ as
\begin{equation}
(\delta\lambda)^2 \geq \frac{\log (2)}{t^2}\frac{k_B T}{  \Delta Q_E^\lambda}.
\end{equation}
This result implies that in order to achieve a better measurement precision, it would either require a longer interrogation time or dissipate more heat for single experimental run. 

To characterize the overall thermodynamic performance of a typical quantum metrology setting, we average the heat dissipation $\Delta Q_E=\langle \Delta Q_E^\lambda\rangle $ over all possible values of the unknown parameter $\lambda$. In the metrological framework based on the interrogation Hamiltonian in Eq.\,\eqref{eq:trad-H}, starting from a pure initial state $|\psi\rangle$, the parametrized final state is given by $\rho_\lambda=\vert\psi_\lambda\rangle\langle \psi_\lambda \vert$. Here $|\psi_\lambda\rangle=e^{-i H_\lambda t}|\psi\rangle$ takes the simple form of a linear superposition $|\psi_\lambda\rangle=\sum_a c_a e^{-i a\lambda t}|a\rangle$ in the eigenbasis defined by $h$ (i.e., $h|a\rangle=a|a\rangle$) with coefficients $c_a$. Note that the QFI $F_Q[\rho_\lambda]\equiv F_Q$ is independent of the parameter $\lambda$ in this scenario \cite{supplement}.  Since the parameter $\lambda$ is unknown, we shall introduce a density matrix to describe 
the ensemble of $\lambda$ parametrized quantum states $\{\rho_\lambda\}$, which is denoted as $\rho_s=\langle\rho_\lambda\rangle$. For a general measurement protocol which is not necessarily optimal, we obtain that \cite{supplement}
\begin{equation}
\label{eq:proof-2}
\mathcal{S}\geq S(\rho_s)\geq \log (2) t^{-2} F_Q.
\end{equation}
Here, $\mathcal{S}$ is the Shannon entropy of the measurement outcomes stored in the memory (see Fig.\,\ref{fig:model}a), and $S(\rho_s)=-\text{tr}(\rho_s\log\rho_s)$ represents the von Neumann entropy of the density matrix $\rho_s$. As $\Delta Q_E \geq k_B T \mathcal{S}$ stated by Landauer's principle, Eq.\,\eqref{eq:proof-2} straightforwardly leads to $\Delta Q_E \geq k_B T S(\rho_s)$, namely
\begin{equation}
  \label{eq:Q-Fq-average}
  \Delta Q_E \geq \log (2) k_B T t^{-2} F_Q.
\end{equation}

{\it Erasure of quantum Fisher information.---}The physical limit on energy consumption in the above quantum metrological framework is derived from the perspective of measurement. Similar to Landauer's principle, which relates the energy cost in computations with the erasure of information, we demonstrate that the heat dissipation in quantum metrology is essentially determined by the erasure of QFI. The measurement process shown in Fig.\,\ref{fig:model}a erases the QFI of the entity including the probe system and the memory, which can be represented by a "many-to-one" map $\mathcal{E}_s: |\psi_\lambda\rangle\langle \psi_\lambda|\otimes |x\rangle\langle x| \to |0\rangle\langle 0|\otimes |x\rangle\langle x|$ for arbitrary $\lambda$. The erasure of QFI can be realized by a map of the more general form $\mathcal{E}_s: |\psi_\lambda\rangle\langle \psi_\lambda| \to \rho$ via the interaction with environment. Such a map transforms the system into a parameter-independent state $\rho$ that contains no information on the unknown parameter $\lambda$. Following Landauer's principle in the quantum regime \cite{Reeb2014} and using the inequality in Eq.\,\eqref{eq:proof-2}, we find that \cite{supplement}
\begin{equation}
     \label{eq:Q-S-F}
    \Delta Q_E+k_B T S(\rho) \geq \log (2) k_B T t^{-2} F_Q,
\end{equation}
where $\Delta Q_E$ represents the heat dissipation induced by the QFI-erasure map $\mathcal{E}_s$ averaged over all possible values of the unknown parameter $\lambda$, and $S(\rho)$ denotes the von Neumann entropy of the parameter-independent state $\rho$. 

The result in Eq.\,\eqref{eq:Q-S-F} implies that the erasure of QFI leads to either heat dissipation into environment (i.e., $\Delta Q_E$), or entropy increase of the system from the pure state $|\psi_\lambda\rangle$ to a possible mixed state $\rho$. One notes that a subsequent transformation of the parameter-independent state $\rho$ into the initial state $\rho_0$ may also dissipate heat according to Landauer's principle. Therefore, the overall heat dissipation will be lower bounded by the right-hand side of Eq.\,\eqref{eq:Q-S-F}, which is consistent with the result in Eq.\,\eqref{eq:Q-Fq-average}. We remark that the thermodynamic bound for the erasure of QFI can be further improved to $\mathcal{Q}_s(\Delta_F)$  \cite{supplement} with $\Delta_F=\log (2)t^{-2} F_Q -S(\rho)$ via the low-temperature correction \cite{Timpanaro2020}. Below we illustrate the behaviors of the bound in Eq.\,\eqref{eq:Q-S-F} using two well-known examples, i.e., the quantum Rabi model and the pure dephasing process.

\begin{figure}
\centering
\includegraphics[width=76mm]{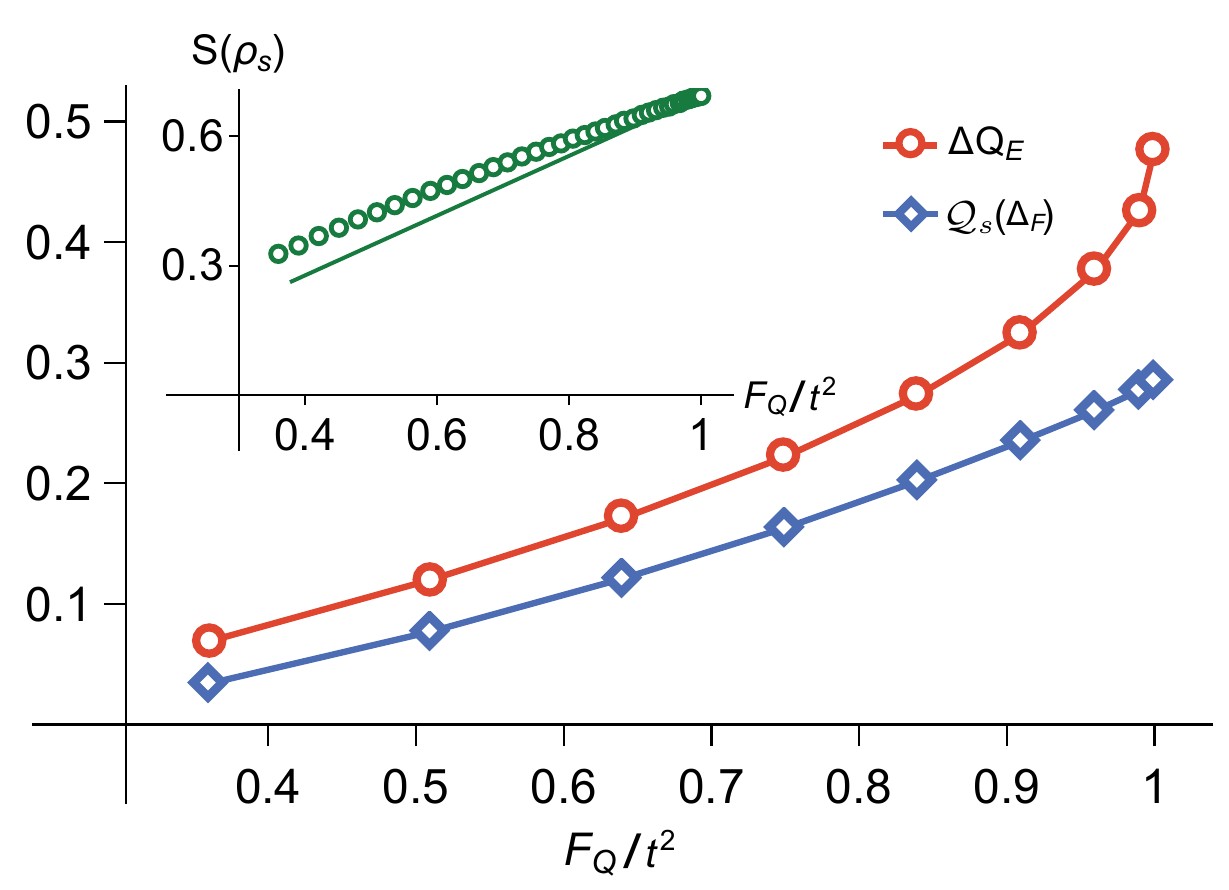}
\caption{\textbf{Erasure of quantum Fisher information via single bosonic mode environment.} Starting from the parametrized states $|\psi_\lambda\rangle=\exp [-i(\lambda \sigma_z/2)  t](c_0|0\rangle+c_1|1\rangle)$ with the QFI $F_Q=4|c_0|^2 |c_1|^2 t^2$, the qubit-probe system approximately evolves into the ground state $|0\rangle$ for different values of $\lambda$ after a certain time interacting with the environment of a bosonic mode. The blue diamonds and the red circles correspond to the thermodynamic bound $\mathcal{Q}_s(\Delta_F)$ and the exact heat dissipation respectively caused by the erasure of QFI. In addition, the inset shows the exact value of the von Neumann entropy $S(\rho_s)$ (circles) and the bound $\log(2)t^{-2} F_Q$ (solid line) given in Eq.\,\eqref{eq:proof-2} for different values of QFI. The parameters are $\omega=\Omega=1$, $g=0.05$, and the erasure time is $\tau_0=30.8$. The bosonic mode is initially in a thermal state at a temperature $T=0.3$.}\label{fig:QRM}
\end{figure}

In the first example of quantum Rabi model, a qubit-probe system in the parametrized state $|\psi_\lambda\rangle$ interacts with  environment, i.e., the bosonic mode initially in a thermal equilibrium state. The global dynamics is governed by the Hamiltonian $H=(\Omega/2)\sigma_z+\omega a^\dagger a- g (a+a^\dagger)\sigma_x$, where $a$ is the bosonic mode and $\sigma_{x,z}$ are Pauli matrices. After a certain evolution time, the qubit probe evolves into an approximately identical final state for different values of the parameter $\lambda$ \cite{supplement}. In Fig.\,\ref{fig:QRM}, we show the heat dissipation into the bosonic mode $\Delta Q_E$ and the corresponding bound from initial states that give rise to different QFI. The results confirm that the heat dissipation is lower bounded by the QFI and the entropy increase of the qubit-probe system.

\begin{figure*}
\centering
\includegraphics[width=16cm]{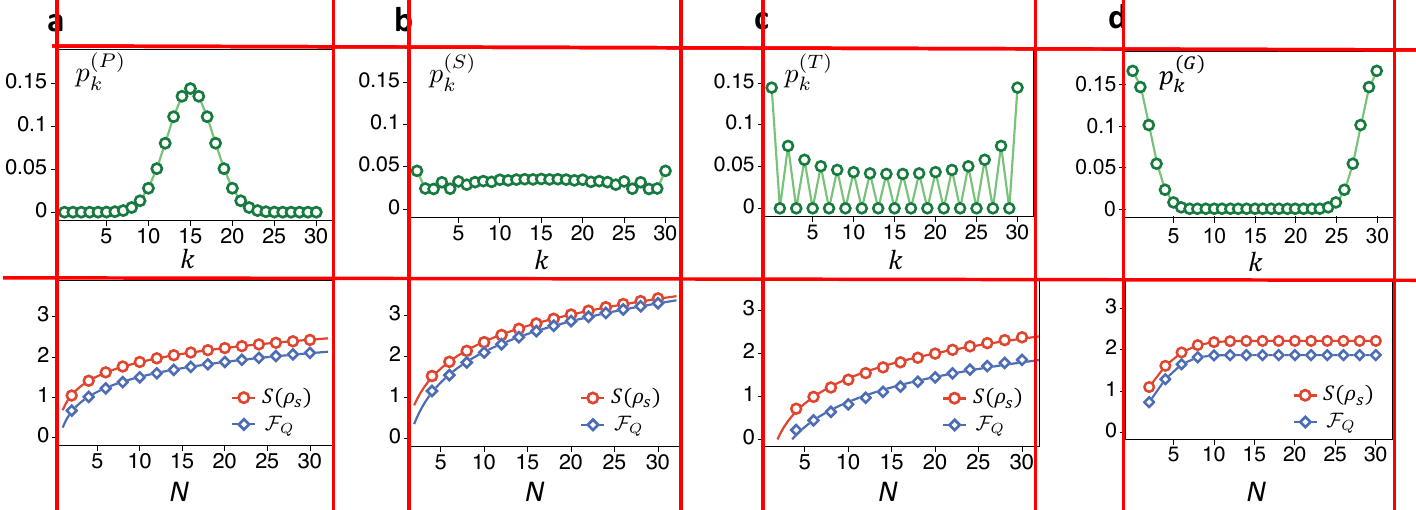}
\caption{\textbf{Von Neumann Entropy of quantum metrology based on symmetric multi-qubit states.} The upper panels depict the probability distribution functions $\{p_k^{(P)},p_k^{(S)},p_k^{(T)},p_k^{(G)}\}$ for four types of symmetric states: product states \textbf{(a)}, squeezed states \textbf{(b)}, twin-Fock states \textbf{(c)}, and GHZ-like states \textbf{(d)}. The lower panels show the corresponding von Neumann entropy $S(\rho_s)$ (red circles), the first three ones of which are well fitted by the logarithmic functions of a form $\sim \alpha\log (N)+\beta$ (solid line) with $\alpha\approx0.51$, $\beta\approx 0.7$ for \textbf{(a)}, $\alpha\approx0.95$, $\beta\approx 0.17$ for \textbf{(b)}, and $\alpha\approx0.84$, $\beta\approx -0.52$ for \textbf{(c)}. In panel \textbf{(d)}, the von Neumann entropy $S(\rho_s)$ saturates to a constant value for a large $N$. In \textbf{(a-d)}, the weighted quantum Fisher information $\mathcal{F}_Q$ (blue diamonds) [Eq.\eqref{eq:S-Fq-multi}] shows the same scaling feature as the von Neumann entropy $S(\rho_s)$. 
}\label{fig:Dicke}
\end{figure*}

In the second example, we consider a spin-$\frac{1}{2}$ probe system coupled with a spin-bath environment via the Ising-type interaction. Under the assumptions of a large spin bath and weak coupling, the system undergoes a pure dephasing process. When the QFI is completely erased, the system evolves from parametrized states $\vert \psi_\lambda\rangle=(1/\sqrt{2})(\vert{\uparrow}\rangle+e^{-i\lambda t}\vert{\downarrow}\rangle)$ to the mixed state $\rho=(1/2)(\vert {\uparrow}\rangle \langle {\uparrow}\vert + \vert{\downarrow}\rangle \langle{\downarrow}\vert)$ for arbitrary $\lambda$. During this process, there is no energy exchange between the probe system and the spin bath, i.e., $\Delta Q_E=0$. The physical consequence arising from the erasure of QFI becomes entropy increase of the probe system, which equals to the bound set by the QFI, namely $S(\rho)=\log(2)t^{-2}F_Q=\log(2)$.

{\it Application to multi-qubit quantum metrology.---}As a unique advantage, quantum metrology based on multi-qubit entangled states can beat the standard quantum limit (SQL) of measurement precision \cite{Giovannetti2011}. We consider symmetric multi-qubit states, which include a wide range of entangled states that can surpass the SQL such as squeezed spin states, twin-Fock states and GHZ states \cite{Pezze2018}. Using the basis of Dicke states $\{|D_N^k\rangle\}_{k=0}^N$  ($N$ is the system size), the symmetric multi-qubit states can be expanded as follows
\begin{equation}
  |\psi\rangle=\sum_{k=0}^N c_k |D_N^k\rangle,
\end{equation}
where $\{p_k=\lvert c_k\rvert^2\}_{k=0}^N$ is the corresponding probability distribution function   \cite{Pezze2018}.

We choose the interrogation Hamiltonian as $H_\lambda=\lambda \sum_j\sigma_j^z/2$, and consider four types of symmetric multi-qubit states, the probability distribution functions of which are shown in Fig.\,\ref{fig:Dicke}\,(a-d). In this scenario, the average heat dissipation is lower bounded as $\Delta Q_E\geq k_B T S(\rho_s)$ [see Eq.\,\eqref{eq:proof-2} and \eqref{eq:Q-Fq-average}], thus we compare the von Neumann entropy $S(\rho_s)$ for these different symmetric multi-qubit states. It can be seen that the scaling of the von Neumann entropy with the system size can be fitted by functions of the form $S(\rho_s)=\alpha \log (N)+\beta$ for product states (a), squeezed states (b), and twin-Fock states (c). Fig.\,\ref{fig:Dicke}\,(d) shows a different type of symmetric multi-qubit states with a probability distribution function $p_k^{(G)} \sim \exp[-k^2/2\nu^2]+\exp[(N-k)^2/2\nu^2]$ (referred to as GHZ-like states below). Particularly, we arrive at the GHZ state in the limit of $\nu\to 0$. It is remarkable that the von Neumann entropy $S(\rho_s)$ for GHZ-like states saturates to a constant value in the large $N$ limit, which further implies that the physical limit on energy consumption of quantum metrology would not increase with the number of qubits involved. Our results demonstrate that, although squeezed spin states, twin-Fock states and GHZ-like states can all beat the SQL of measurement precision \cite{supplement}, their energy efficiencies in quantum metrology are radically different. The GHZ-like states represent a class of resource states for quantum metrology that can offer superior advantages not only in measurement precision but also in energy consumption.

Such an interesting phenomenon of energy consumption in multi-qubit quantum metrology results from the fact that the von Neumann entropy $S(\rho_s)$ is related to a weighted summation of the QFI, $\mathcal{F}_Q$, contributed by all level pairs, namely
\begin{equation}
  \label{eq:S-Fq-multi}
  S(\rho_s)\geq \mathcal{F}_Q=\frac{1}{t^2}\sum_{a<b} \frac{F_Q^{ab}}{2(a-b)^2}\log\left(\frac{2}{p_a+p_b}\right),
\end{equation}
where $F_Q^{ab}$ represents the QFI contributed by the level pair $(a,b)$, the direct summation of which equals to the total QFI $F_Q$ \cite{supplement}. 
It can be seen from Fig.\,\ref{fig:Dicke} that $\mathcal{F}_Q$ provides a well-behaved bound for  $S(\rho_s)$, and reveals its scaling feature. Therefore, by focusing on the weighted QFI $\mathcal{F}_Q$, we are able to analyze energy consumption and engineer suitable quantum resource states in order to reduce heat dissipation while sustaining high measurement precision.


{\it Summary \& outlook.---} The work establishes a fundamental link between the concepts of entropy, QFI and heat dissipation, and clearly states the physical limit on energy consumption of quantum metrology to achieve a certain level of measurement precision. These results are only relevant in the asymptotic and local regimes of quantum estimation theory. It would be interesting to extend them into more realistic scenarios with prior information beyond the quantum CRB \cite{Kay1993,Personick1971,Macieszczak2014,Demkowicz2015,Demkowicz2020,Rubio2020,Lee2022,Sidhu2020}.
The revealed thermodynamic principle for quantum metrology provides a new perspective to explore quantum advantage, apart from the considerable focusing on the measurement precision, offered by quantum resources. We show that multi-qubit states that can achieve similar performances in measurement precision may demonstrate very different energy efficiencies. 
Besides, the QFI, equivalent to quantum metric, can characterize the property of a given band defined over parameter space. Therefore, the present connection between quantum Fisher information and thermodynamic quantities may offer a new way to explore the topological properties of energy bands from a thermodynamic perspective.
{\it Acknowledgments.---} We thank Martin B. Plenio, Nathan Goldman and Zhendong Zhang for helpful discussions and suggestions. The work is supported by National Natural Science Foundation of China (Grant No.~12161141011, 11874024, 11690032), the National Key R$\&$D Program of China (Grant No. 2018YFA0306600), the Open Project Program of Wuhan National Laboratory for Optoelectronics (2019WNLOKF002), the Fundamental Research Funds for the Central Universities, and the Interdisciplinary program of Wuhan National High Magnetic Field Center (Grand No. WHMFC202106).

\bibliography{reference}
 

\section*{Supplementary material}

# Supplementary Materials for

## Thermodynamic principle for quantum metrology

Yaoming Chu and Jianming Cai

Corresponding author: jianmingcai@hust.edu.cn

**The PDF file includes:**

Supplementary Methods

Fig. S1 and S2

References and Notes

## Supplementary Methods

### Theoretical description of measurement

We start from a general pure parametrized state $|\psi_\lambda\rangle$ of the metrological system, the typical measurement operation on which can be described as an interaction with the measurement apparatus (memory),

$$H_{\text{int}}^{SM} = g\hat{M} \otimes \hat{P} \qquad (S.1)$$

where $\hat{M}$ is the system observable associated with the measurement basis $\{|m\rangle\langle m|\}_{m=1}^{d}$ in the main text, and $\hat{P}$ denotes the apparatus operator. After a suitable interrogation time governed by $H_{\text{int}}^{SM}$, the combined metrology-apparatus system can evolve into the following entangled state[1]

$$|\Psi^{SM}\rangle = \sum_m c_m^\lambda |m\rangle \otimes |x_m\rangle, \qquad (S.2)$$

where $\{|x_m\rangle\}$ corresponds to the pointer basis of the apparatus. Such an entangled state is quickly decohered into a mixed state diagonal in the pointer basis due to the interaction between the apparatus and its surrounding environment [1, 2]

$$\rho^{SM} = \sum_m \alpha_m^\lambda |m\rangle\langle m| \otimes |x_m\rangle\langle x_m|, \quad \alpha_m^\lambda = |c_m^\lambda|^2. \qquad (S.3)$$

Thus, we arrive at Eq. (1) in the main text.

### Proof of Theorem 1

Before proving Theorem 1, we first introduce the following lemma.

**Lemma 1** The function $f(p) = -p \log p + 2 \log(2)p^2$ is subadditive for $p \in [0, 1]$, i.e.

for $p_1, p_2, p_1 + p_2 \in [0, 1]$,

$$f(p_1 + p_2) \leq f(p_1) + f(p_2). \tag{S.4}$$

*Proof of Lemma 1* We first note that the function $g(q) = 4 \log(2) q(1-q) + q \log(q) + (1-q) \log(1-q)$ is non-positive for $q \in [0, 1]$. Thus, we can obtain

$$4 \log(2) q_1 q_2 + q_1 \log(q_1) + q_2 \log q_2 \leq 0 \tag{S.5}$$

for $q_1, q_2 \in [0, 1]$ and $q_1 + q_2 = 1$. Setting $q_1 = p_1/(p_1 + p_2)$ and $q_2 = p_2/(p_1 + p_2)$, this inequality leads to that

$$4 \log(2) \left( \frac{p_1 p_2}{p_1 + p_2} \right) + p_1 \log \left( \frac{p_1}{p_1 + p_2} \right) + p_2 \log \left( \frac{p_2}{p_1 + p_2} \right) \leq 0. \tag{S.6}$$

By using $p_1 + p_2 \leq 1$ and Eq. (S.6), we get

$$f(p_1 + p_2) - f(p_1) - f(p_2) = 4 \log(2) p_1 p_2 + p_1 \log \left( \frac{p_1}{p_1 + p_2} \right) + p_2 \log \left( \frac{p_1}{p_1 + p_2} \right) \leq 0. \tag{S.7}$$

Thus, Lemma 1 can be directly proved.

In quantum metrology, the optimal measurement basis to achieve the best measurement precision corresponds to the eigenbasis of the symmetric logarithmic derivative $L_\lambda$, which is denoted as $\{|\ell_\lambda\rangle\langle\ell_\lambda|\}_{\ell=1}^d$ [3]. The probability distribution function of the parametrized final state $\rho_\lambda = |\psi_\lambda\rangle\langle\psi_\lambda|$ in this basis is given by $p_\ell^\lambda = |\langle\ell_\lambda|\psi_\lambda\rangle|^2$. Thus, the Shannon entropy of the measurement outcomes can be expressed as

$$\mathcal{S} = -\sum_\ell p_\ell^\lambda \log(p_\ell^\lambda). \tag{S.8}$$

The QFI of $\rho_\lambda$ with respect to the unknown parameter $\lambda$ can be related to the variance of $L_\lambda$ as [3]

$$F_Q[\rho_\lambda] = \mathrm{Var}[L_\lambda, \rho_\lambda] = \frac{1}{2}\sum_{\ell,\ell'}(\ell_\lambda - \ell'_\lambda)^2 p_\ell^\lambda p_{\ell'}^\lambda, \tag{S.9}$$

where $\ell_\lambda$ is the eigenvalue of $L_\lambda$ associated with the eigenvector $|\ell_\lambda\rangle$. Hence, we have

$$2F_Q[\rho_\lambda]/\|L_\lambda\|^2 \le \sum_{k\neq\ell} p_k^\lambda p_\ell^\lambda = 1 - \sum_\ell (p_\ell^\lambda)^2, \tag{S.10}$$

where $\|L_\lambda\| \le 2\|h_\lambda\|$ according to $L_\lambda = 2i[\rho_\lambda, h_\lambda]$. In addition, the subadditivity of the function $f(p) = -p\log p + 2\log(2)p^2$ in Lemma 1 straightforwardly leads to the following inequality for $\sum_\ell p_\ell = 1$,

$$-\sum_\ell p_\ell \log p_\ell + 2\log 2 \sum_\ell p_\ell^2 \ge f(1) = 2\log 2. \tag{S.11}$$

Therefore, we can prove Eq. (2) in the main text, namely

$$\mathcal{S} \ge \log(2)\|h_\lambda\|^{-2} F_Q[\rho_\lambda]. \tag{S.12}$$

Theorem 1 can be proved by invoking Landauer's principle $\Delta Q_E^\lambda \ge k_B T \mathcal{S}$ in combination with Eq. (S.12).

**Average heat dissipation in quantum metrology**

In the traditional quantum metrological framework, the interrogation Hamiltonian can be written as $H_\lambda = \lambda h$, where $\lambda$ is the unknown parameter to be estimated. After the interrogation, the QFI of the parametrized final state $\rho_\lambda = |\psi_\lambda\rangle\langle\psi_\lambda|$ with $|\psi_\lambda\rangle = e^{-iH_\lambda t}|\psi\rangle = \sum_a c_a e^{-ia\lambda t}|a\rangle$ is given by

$$F_Q = 2t^2 \sum_{a,b}(a-b)^2 p_a p_b, \tag{S.13}$$

3

where $p_a = |c_a|^2$. Without loss of generality, we can assume that the eigenbasis of $h$ is nondegenerate. If this is not case, for example, $\{|a_1\rangle, \cdots, |a_r\rangle\}$ are eigenvectors associated with the same eigenvalue $a$, we can define a new basis state $|a\rangle \sim \sum_{i=1}^{r} \langle a_i|\psi\rangle |a_i\rangle$ up to a normalization factor. In this way, the density matrix $\rho_s = \langle \rho_\lambda \rangle$ can be explicitly written as

$$\rho_s \propto \sum_{a,b} c_a c_b^* |a\rangle \langle b| \int e^{-i(a-b)\lambda t} \mathrm{d}\lambda. \tag{S.14}$$

After averaging over all possible values of $\lambda$, $\rho_s$ has a diagonal form of

$$\rho_s = \sum_a p_a |a\rangle \langle a|. \tag{S.15}$$

The corresponding von Neumann entropy is

$$S(\rho_s) = -\sum_a p_a \log p_a. \tag{S.16}$$

Based on the subsidiary inequality in Eq. (S.11), $S(\rho_s)$ is found to satisfy

$$S(\rho_s) \geq 2\log(2) \left(1 - \sum_a p_a^2\right) = 2\log(2) \sum_{a \neq b} p_a p_b. \tag{S.17}$$

For a normalized energy scale (i.e. the operator seminorm $\|h\| = 1$), we have $(a-b)^2 \leq 1$. Thus, the QFI in Eq. (S.13) is constrained by $t^{-2} F_Q \leq 2 \sum_{a \neq b} p_a p_b$, which further leads to that [namely Eq. (6) in the main text]

$$S(\rho_s) \geq \log(2) t^{-2} F_Q \tag{S.18}$$

On the other hand, the Shannon entropy associated with the measurement outcomes in a general basis of $\{|m\rangle \langle m|\}_{m=1}^{d}$ can be expressed as

$$\mathcal{S} = \sum_m f_s(\alpha_m) \tag{S.19}$$

4

with $f_s(p) = -p \log p$ and $\alpha_m = \langle \alpha_m^\lambda \rangle = \langle m | \rho_s | m \rangle$. Following the concavity of $f_s(p)$, and using $\alpha_m = \sum_a p_{m|a} p_a$ with $p_{m|a} = |\langle m | a \rangle|^2$, we obtain that

$$\mathcal{S} \geq \sum_a \sum_m p_{m|a} f(p_a) = \sum_a f(p_a) = S(\rho_s). \tag{S.20}$$

According to Landauer's principle, we have $\Delta Q_E \geq k_B T \mathcal{S} \geq k_B T S(\rho_s)$, and thus the thermodynamic bound of quantum metrology with general measurements [see Eq. (7)] can be proved.

**Heat dissipation during the erasure of QFI**

A general QFI erasure channel of the system $S$ can be realized via interacting with a thermal environment $E$ by means of a global unitary $U$ [4, 5], namely

$$\mathcal{E}_s(\rho_\lambda) = \mathrm{tr}_E[U(\rho_\lambda \otimes \varrho_E^{\mathrm{th}})U^\dagger] = \rho \tag{S.21}$$

for arbitrary $\lambda$. Here, we assume the environment is in a thermal equilibrium state of the form

$$\varrho_E^{\mathrm{th}} \sim e^{-H_E/k_B T}, \tag{S.22}$$

where $H_E$ is the Hamiltonian of the environment and $T$ is the temperature. In such a scenario, the heat dissipation is defined as the energy increase of the environment [4]

$$\Delta Q_E^\lambda = \mathrm{tr}[H_E(\varrho_E^\lambda - \varrho_E^{\mathrm{th}})], \tag{S.23}$$

where $\varrho_E^\lambda = \mathrm{tr}_S[U(\rho_\lambda \otimes \varrho_E^{\mathrm{th}})U^\dagger]$. By averaging over the parameter $\lambda$, we obtain the average heat dissipation as follows

$$\Delta Q_E = \langle \Delta Q_E^\lambda \rangle = \mathrm{tr}\left[H_E\left(\langle \varrho_E^\lambda \rangle - \varrho_E^{\mathrm{th}}\right)\right] \tag{S.24}$$

5

with $\langle \varrho_E^\lambda \rangle = \mathrm{tr}_S[U(\rho_s \otimes \varrho_E^{\mathrm{th}})U^\dagger]$. Thus, one can interpret $\Delta Q_E$ as the heat generation of the system evolving from the state $\rho_s$ to the final parameter-independent state $\rho$ by interacting with a thermal environment, which straightforwardly leads to that

$$\Delta Q_E \geq k_B T[S(\rho_s) - S(\rho)], \tag{S.25}$$

according to Landauer's principle. Using $S(\rho_s) \geq \log(2)t^{-2}F_Q[\rho_\lambda]$ in Eq. (S.17), we can prove the inequality regarding the erasure of QFI in Eq. (8) in the main text, namely

$$\Delta Q_E + k_B T S(\rho) \geq \log(2)t^{-2} k_B T F_Q. \tag{S.26}$$

**Improved bound for the erasure of QFI**

Based on the corrected Landauer's principle at low temperatures in Ref. [5], the inequality in Eq. (S.25) can be improved to

$$\Delta Q_E \geq \mathcal{Q}_S \left( S(\rho_s) - S(\rho) \right), \tag{S.27}$$

where $\mathcal{Q}_s = f_Q \circ f_S^{-1}$ is a function depending on the details of the environment [5]. The $\mathcal{Q}_s$ is a monotonically increasing function, which allows us to obtain

$$\Delta Q_E \geq \mathcal{Q}_S(\Delta_F), \tag{S.28}$$

where

$$\Delta_F = \log(2)t^{-2}F_Q - S(\rho). \tag{S.29}$$

In the quantum Rabi model, the bosonic mode is described by the Hamiltonian $H_E = \omega a^\dagger a$, one can find that

$$
\begin{aligned}
f_Q(\beta') &= f_E(\beta') - f_E(\beta), \\
f_S(\beta') &= \beta' f_E(\beta') - \beta f_E(\beta) + f_Z(\beta') - f_Z(\beta),
\end{aligned}
\tag{S.30}
$$

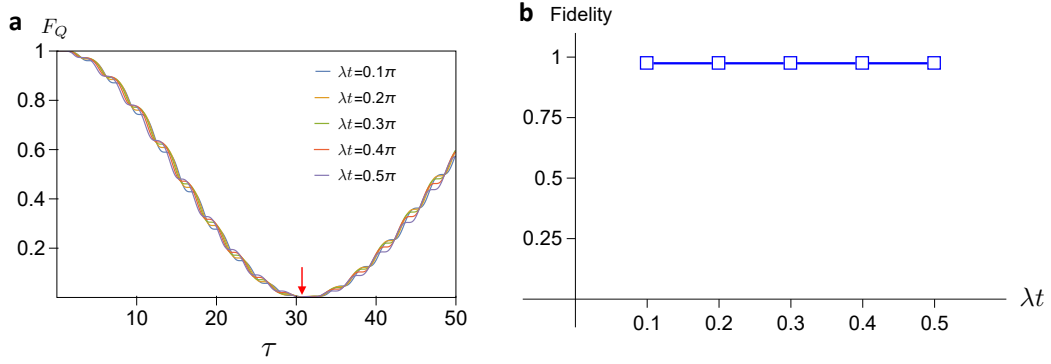

Figure S1: **Erasure of quantum Fisher information via single bosonic mode environment in quantum Rabi model.** The qubit-probe system is initially in the parametrized states $|\psi_\lambda\rangle = \exp[-i(\lambda\sigma_z/2)t](|0\rangle + |1\rangle)/\sqrt{2}$ with the QFI $F_Q = t^2$ and $t$ the interrogation time. The bosonic mode environment is assumed to be in a thermal state of temperature $T$. Their global dynamics is governed by the Hamiltonian $H = (\Omega/2)\sigma_z + \omega a^\dagger a - g(a + a^\dagger)\sigma_x$, where $a$ is the bosonic mode and $\sigma_{x,z}$ are Pauli matrices. **a,** The QFI of the probe system is gradually erased via interacting with the bosonic mode. The red arrow denotes the almost complete erasing point for different values of $\lambda$ with a same evolution time $\tau_0 \approx 30.8$. **b,** After the fixed evolution time $\tau_0$, we obtain the fidelity (blue squares) between the final state and the state $|0\rangle$ of the probe system for different values of $\lambda$. It can be seen that the fidelity is very close to 1, which indicates that the qubit-probe system approximately evolves into the same state $|0\rangle$ for different values of $\lambda$. We remark that this feature holds as well for more general parametrized states $|\psi_\lambda\rangle = \exp[-i(\lambda\sigma_z/2)t](c_0|0\rangle + c_1|1\rangle)$ with the QFI $F_Q = 4|c_0|^2|c_1|^2t^2$. The parameters are set as $\omega = \Omega = 1$, $g = 0.05$, and $T = 0.3$.

where $\beta = 1/T$ and $T$ is the initial temperature of the environment. Here, the functions $f_E(\beta')$ and $f_Z(\beta')$ are defined as

$$
\begin{aligned}
f_E(\beta') &= \frac{\omega}{e^{\beta'\omega} - 1}, \\
f_Z(\beta') &= -\log(1 - e^{-\beta'\omega}).
\end{aligned}
\tag{S.31}
$$

**Quantum metrology using squeezed spin states**

We consider spin squeezed states, which can be generated by one-axis twisting dynamical evolution along the $\hat{z}$ axis [6], namely

$$
|\psi_{\text{OAT}}\rangle = e^{-i\frac{u}{2}J_z^2}\left|\frac{\pi}{2}, 0, N\right\rangle,
\tag{S.32}
$$

where $|\frac{\pi}{2}, 0, N\rangle$ is the coherent spin state pointing along the positive $\hat{x}$ axis, and $J_n$ ($n =$

$x, y, z$) represent the collective spin operators for $N$ spins, i.e. $J_n = \sum_{j=1}^{N} \sigma_j^n / 2$. The squeezed and antisqueezed spin components of the state $|\psi_{\text{OAT}}\rangle$ is given by

$$J_s = J_z \cos(\delta) - J_y \sin(\delta),$$
$$J_{\text{as}} = J_y \cos(\delta) + J_y \sin(\delta) \tag{S.33}$$

with $\delta = (1/2) \arctan(B/A)$ the squeezing angle in terms of $A = 1 - \cos^{N-2}(u)$ and $B = 4 \sin(u/2) \cos^{N-2}(u/2)$ [6, 7]. We choose $u = 2.4/N^{2/3}$ to achieve the maximal squeezing [8].

In the quantum metrological framework with the interrogation Hamiltonian $H_\lambda = \lambda J_z$, in order to maximize the QFI, we further perform a unitary rotation along the $\hat{x}$ axis on the squeezed state $|\psi_{\text{OAT}}\rangle$ to prepare the initial state

$$|\psi\rangle = e^{-i(\frac{\pi}{2} - \delta) J_x} |\psi_{\text{OAT}}\rangle. \tag{S.34}$$

Based on Eq. (S.32) and Eq. (S.34), we can obtain the probability distribution function $\{p_k^{(S)}\}$ of the initial state $|\psi\rangle$ in the Dicke basis [7] and further the von Neumann entropy $S(\rho_s) = -\sum_{k=1}^{d} p_k^{(S)} \log(p_k^{(S)})$, see Fig. 3(b) in the main text. In addition, we find that the QFI of $|\psi_\lambda\rangle = e^{-i\lambda t J_z} |\psi\rangle$ with respect to the unknown parameter $\lambda$ scales as $F_Q \sim N^{1.7} t^2$ in the large $N$ limit [7], which demonstrates that quantum metrology using squeezed spin states can surpass the standard quantum limit of measurement precision.

**The probability distribution function of twin-Fock states**

Twin-Fock states represent an interesting type of Dicke states, which provide the maximal QFI among all Dicke states [7]. For quantum metrology based on the interrogation Hamiltonian $H_\lambda = \lambda J_z$ (as we consider in the main text), we choose twin-Fock state along the $\hat{x}$

axis, i.e. $|0_x\rangle = e^{-i\pi J_y/2}|0_z\rangle = e^{-i\pi J_y/2}|D_N^{N/2}\rangle$, where $|D_N^{N/2}\rangle$ denotes the Dicke state with $N/2$ excitations along $\hat{z}$ the axis [9]. The corresponding probability distribution function can be written as follows

$$p_k^{(T)} = \begin{cases} \dfrac{C_N^{N/2}}{C_N^k}\dfrac{1}{2^N}\left[\displaystyle\sum_{m=0}^{k} C_{N/2}^m C_{N/2}^{k-m}(-1)^{N/2-k+m}\right]^2 & \text{for} \quad k \le N/2, \\[4mm] \dfrac{C_N^{N/2}}{C_N^k}\dfrac{1}{2^N}\left[\displaystyle\sum_{m=0}^{N-k} C_{N/2}^m C_{N/2}^{N-k-m}(-1)^{N-k-m}\right]^2 & \text{for} \quad k > N/2. \end{cases} \tag{S.35}$$

This analytical formula allows us to efficiently calculate the von Neumann entropy $S(\rho_s) = -\sum_{k=1}^{d} p_k^{(T)} \log(p_k^{(T)})$ and obtain the physical limit on energy consumption of quantum metrology using twin-Fock states. The QFI of $|\psi_\lambda\rangle = e^{-i\lambda t J_z}|0_x\rangle$ with respect to the unknown parameter $\lambda$ is given by $F_Q = N(N+2)/2$ [7], which implies that quantum metrology using twin-Fock states can reach the Heisenberg limit of measurement precision.

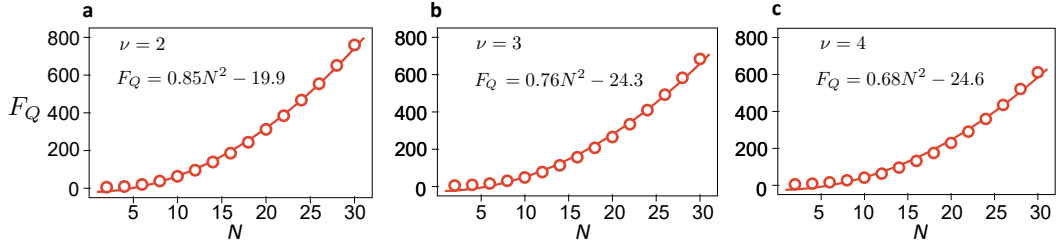

Figure S2: **Quantum Fisher information for GHZ-like states as a function of the system size.** It can be seen that the QFI is well fitted by $F_Q = \alpha N^2 + \beta$ for different values of $\nu$ with $\alpha = 0.85$ and $\beta = -19.9$ **(a)**, $\alpha = 0.76$ and $\beta = -24.39$ **(b)**, $\alpha = 0.68$ and $\beta = -24.6$ **(c)**.

**Scaling of the QFI for quantum metrology using GHZ-like state**

In the main text, we find that the physical limit on energy consumption of quantum metrology using certain types of multi-qubit entangled states, e.g. GHZ-like states, would not

increase with the number of qubits involved. The probability distribution function of such a type of resource state in the Dicke basis is described as

$$p_k^{(G)} \sim \exp[-k^2/2\nu^2] + \exp[(N-k)^2/2\nu^2] \tag{S.36}$$

In Fig. S2, we show the scaling of the QFI $F_Q$ with the system size $N$ for different values of $\nu$. In can be seen that the QFI for GHZ-like states scales as $F_Q \sim N^2$, thus achieving Heisenberg limit of measurement precision.

**Bound for heat dissipation via the weighted QFI**

Here, we provide a brief proof of Eq. (10) in the main text, which is straightforward by rewriting the von Neumann entropy for $\rho_s = \sum_a p_a |a\rangle\langle a|$ [see Eq. (S.15)] as

$$\begin{aligned}
S(\rho_s) &= -\sum_a p_a \log(p_a) \\
&= -\frac{1}{2}\sum_{a,b} p_a p_b \log(p_a) - \frac{1}{2}\sum_{a,b} p_a p_b \log(p_b) \\
&= -\frac{1}{2}\sum_{a,b} p_a p_b \log(p_a p_b) \\
&= \frac{1}{2}S(\rho_s^2) - \frac{1}{2}\sum_{a\neq b} p_a p_b \log(p_a p_b),
\end{aligned} \tag{S.37}$$

where $S(\rho_s^2) = -\sum_a p_a^2 \log(p_a^2)$. By introducing $R = -\mathrm{tr}\log\rho_s$ and using Titu's lemma, we can prove that

$$\frac{1}{2}S(\rho_s^2) \geq \frac{1}{R}S^2(\rho_s). \tag{S.38}$$

We define the weighted QFI as

$$\mathcal{F}_Q = \frac{1}{t^2}\sum_{a<b} \frac{F_Q^{ab}}{2(a-b)^2} \log\left(\frac{2}{p_a+p_b}\right), \tag{S.39}$$

where $F_Q^{ab} = 4t^2(a-b)^2 p_a p_b$ and the summation $\sum_{a<b} F_Q^{ab} = F_Q$. Exploiting the inequal-

ities $p_a + p_b \geq 2\sqrt{p_a p_b}$, we obtain

$$\mathcal{F}_Q \leq -\frac{1}{2} \sum_{a \neq b} p_a p_b \log(p_a p_b). \tag{S.40}$$

Eq. (S.37), (S.38) and (S.40) together lead to the following inequality

$$\frac{1}{R} S^2(\rho_s) - S(\rho_s) + \mathcal{F}_Q \leq 0, \tag{S.41}$$

which further results in

$$S(\rho_s) \geq \frac{R}{2} \left[ 1 - \left( 1 - \frac{4\mathcal{F}_Q}{R} \right)^{\frac{1}{2}} \right]. \tag{S.42}$$

For a large rank of $\rho_s$, we have $R \gg S(\rho_s) \geq \mathcal{F}_Q$. Therefore, we expand the square root

term in Eq. (S.42) with respect to the small quantity $\mathcal{F}_Q/R$, and neglect higher-order terms

$O(\mathcal{F}_Q^n/R^n)$ with $n \geq 2$, which allows us to obtain [see Eq. (10) in the main text]

$$S(\rho_s) \geq \frac{R}{2} \left[ 1 - \left( 1 - \frac{2\mathcal{F}_Q}{R} \right) \right] = \mathcal{F}_Q. \tag{S.43}$$


 
\end{document}